\documentclass[acmsmall]{acmart}
\AtBeginDocument{%
  }

\setcopyright{cc}
\setcctype{by}
\acmDOI{10.1145/3808191}
\acmYear{2026}
\acmJournal{PACMSE}
\acmVolume{3}
\acmNumber{FSE}
\acmArticle{FSE184}
\acmMonth{7}
\acmSubmissionID{fse26mainb-p1979-p}
\received{2026-02-24}
\received[accepted]{2026-03-24}

\usepackage{geometry}
\usepackage{booktabs}
\usepackage{multirow}
\usepackage{graphicx}
\usepackage{pifont}
\usepackage{array}
\usepackage{makecell}

\newcommand{\cmark}{\ding{51}}%
\newcommand{\xmark}{\ding{55}}%

\newcolumntype{M}[1]{>{\centering\arraybackslash}m{#1}}
\newcolumntype{L}[1]{>{\raggedright\arraybackslash}m{#1}}
\usepackage{url}
\usepackage{tabularx}
\usepackage{algorithmic}
\usepackage{multirow}
\usepackage{graphicx}
\usepackage{textcomp}
\usepackage{xcolor}
\usepackage[notransparent]{svg}
\usepackage{booktabs}
\usepackage{enumitem}
\usepackage{booktabs}
\usepackage{pgfplots}
\usepackage{amsfonts} 
\usepackage{hyperref}
\usepackage{subcaption}
\usepackage{threeparttable}
\usepackage{titlesec}

\titlespacing*{\paragraph}
{0pt}    
{0.3em}    
{0.3em}  

\usepackage{framed}
\usepackage{wrapfig}
\usepackage{tcolorbox}

\newcommand{\github}[1]{GitHub{#1}}

\newcommand{\jira}[1]{Jira{#1}}

\newcommand{\ga}{LinkAnchor}
\newcommand{\rev}[1]{{\color{black}#1}}

\NewDocumentCommand\checkNum{+m}{{\color{black} #1}}
\NewDocumentCommand\code{+m}{{\small\ttfamily  #1}}
\NewDocumentCommand\tool{+m}{{\small\scshape  #1}}
\NewDocumentCommand\citeTodo{+m}{{\color{black}[?]}}

\newcommand{\change}[1]{{\color{black}#1}}

\newcommand{\secref}[1]{Section~\ref{sec:#1}}
\newcommand{\tabref}[1]{Table~\ref{tab:#1}}

\newcommand{\figref}[1]{Figure~\ref{fig:#1}}

\begin{document}
\title{LinkAnchor: An Autonomous LLM-Based Agent for Issue-to-Commit Link Recovery}

\author{Arshia Akhavan}
\orcid{0009-0007-9631-6964}
\affiliation{%
  \institution{Bowling Green State University}
  \department{Department of Computer Science}
  \city{Bowling Green}
  \country{USA}
}
\email{arshiaa@bgsu.edu}

\author{Alireza Hoseinpour}
\orcid{0009-0008-1619-0052}
\affiliation{%
  \institution{Bowling Green State University}
  \department{Department of Computer Science}
  \city{Bowling Green}
  \country{USA}
}
\email{ahosein@bgsu.edu}

\author{Abbas Heydarnoori}
\orcid{0000-0001-9785-2880}
\affiliation{%
  \institution{Bowling Green State University}
  \department{Department of Computer Science}
  \city{Bowling Green}
  \country{USA}
}
\email{aheydar@bgsu.edu}

\author{Hamid Bagheri}
\orcid{0000-0001-6686-466X}
\affiliation{%
  \institution{University of Nebraska-Lincoln}
  \department{School of Computing}
  \city{Lincoln}
  \country{USA}
}
\email{bagheri@unl.edu}

\author{Mehdi Keshani}
\orcid{0000-0002-8647-0067}
\affiliation{%
  \institution{University of Zurich}
  \department{Department of Informatics}
  \city{Zurich}
  \country{Switzerland}
}
\email{mehdi.keshani@uzh.ch}



\begin{abstract}

Issue-to-commit link recovery in software repositories is fundamental to software traceability and project management, yet it remains a challenging task. Prior studies show that only about 42.2\% of issues on GitHub are correctly linked to their commits, highlighting the need for more effective solutions. Existing work has explored a range of ML/DL approaches, and more recently, large language models (LLMs) have been applied to this problem. However, these methods face two major limitations. First, LLMs are restricted by limited context windows and cannot simultaneously process all available data sources, such as long commit histories, extensive issue discussions, and large code repositories. Second, most approaches operate on individual issue--commit pairs, where a model independently scores the relevance of a single commit to an issue. This pairwise formulation fails to account for the complex associativity of software fixes, where an issue is often resolved by an aggregate chain of commits rather than a single atomic change. By ignoring these temporal and parental dependencies, existing methods often fail to incorporate the complete resolution logic and might misidentify intermediate commits as final fixes. Furthermore, this strategy is computationally inefficient in large repositories, as it requires exhaustively evaluating an enormous number of candidate pairs. To address these challenges, we present \ga{}, the first autonomous LLM-based agent designed specifically for issue-to-commit link recovery. \ga{} introduces a lazy-access architecture that allows the underlying LLM to dynamically retrieve only the most relevant contextual data, such as commits, issue comments, and code files, without exceeding token limits. Instead of isolated scoring, \ga{} treats link recovery as a dynamic search process, navigating the commit graph to identify the final resolving commit and effectively aggregating the entire chain of contributing changes. \ga{} is first to formalize ILR as dynamic heuristic search over commit chains (vs. prior pairwise scoring), enabling aggregate reasoning that recovers distributed fixes (46\% of cases). Our evaluations show that \ga{} outperforms state-of-the-art baselines by 41–714\% in Hit@1 across six large-scale open-source projects, while costing only about 0.01 US dollars per issue. Finally, \ga{} is designed and tested for both GitHub and Jira, and its modular architecture makes it straightforward to extend to other platforms.

\end{abstract}





\maketitle

\section{Introduction}\label{sec:intro}

Software traceability concerns identifying and leveraging relationships among software artifacts and other development resources, and such relationships support a broad range of software engineering tasks~\cite{heydarnoori2009supporting,heydarnoori2011two,gharehyazie2019cross,moran2020improving,tahmooresi2020studying,rodriguez2021leveraging,abedini2024can,fuchss2025lissa,khosravani2026lia}. A key problem in this domain is issue-to-commit link recovery (ILR), which seeks to associate issue reports with the corresponding commits in software repositories that resolve them~\cite{kondo2022empirical}. Prior research has shown that ILR is critical for a wide range of tasks, including bug prediction~\cite{ruan2019deeplink,le2015rclinker}, feature location~\cite{dit2013feature}, and security risk assessment~\cite{nguyen2022vulcurator,nguyen2022hermes}. More broadly, ILR also supports project management~\cite{panis2010successful} and the evaluation of software maintenance quality~\cite{sun2017frlink}. Although referencing the related issue in the corresponding commit is considered a good practice, studies show that many issue-commit links are often missing in large-scale software projects.~\cite{ruan2019deeplink,romo2014filling}. Manually recovering these links is a time-consuming and error-prone task for project maintainers and contributors~\cite{mazrae2021automated,ruan2019deeplink}.


\rev{To resolve this problem, researchers have developed a variety of automated techniques for ILR~\cite{mazrae2021automated,nguyen2012multi,ruan2019deeplink,sun2017improving,sun2017frlink,xie2019deeplink,zhang2023ealink,huang2025back,wang2025mplinker,deng2024promptlink}. State-of-the-art approaches primarily frame ILR as either a ranking or a classification task. Ranking systems, like~\cite{zhang2023ealink,huang2025back,mazrae2021automated,ruan2019deeplink,sun2017frlink,xie2019deeplink}, score commits based on how well their text semantically matches the issue's title and description, whereas classification models, such as~\cite{wang2025mplinker,deng2024promptlink}, use the same textual features to assign a binary label (i.e., true-link or false-link) to issue-commit pairs. However, both ILR paradigms rely on individual issue-commit pair analysis, where the model evaluates the relationship between a single issue and a single commit in isolation. While this setup has produced promising results, existing work remains subject to important limitations:}

\begin{enumerate}

\item \emph{L1: Single Commit Evaluation Scheme:}
In real-world settings, issues are often resolved not by a single commit, but by a sequence of related commits on a dedicated branch~\cite{zhang2023ealink,huang2025back}. This distinction is critical: when multiple commits contribute to resolving an issue, no single intermediate commit contains enough context to represent the complete resolution. Moreover, ILR is also used to derive time-related metrics, such as issue resolution duration~\cite{panis2010successful}. In such cases, it is essential to identify the entire resolving commit chain, particularly the first commit that starts the fix and the last commit that completes it, rather than treating the resolution as a single isolated event. Baseline methods, however, are designed to evaluate and score commits in isolation~\cite{huang2025back,wang2025mplinker,deng2024promptlink,xie2019deeplink,ruan2019deeplink,mazrae2021automated}. As a result, they cannot capture the \change{semantic and chronological} relationships between commits within a chain, and thus fail to reconstruct the full picture of how the issue was resolved.

\item \emph{L2: Limited Utilization of Available Context:}
Most existing studies fail to fully exploit all available data and context for ILR, since incorporating every contextual source, such as discussion threads or the complete repository codebase, can inflate training datasets, which introduces noise, additional parameters, and overfitting risks~\cite{zhang2023ealink}. In the case of LLMs, it can also quickly exceed the context window~\cite{huang2025back}. To mitigate these challenges, many approaches neglect to incorporate crucial but lengthy contextual data. In practice, this selective sampling and feature engineering can omit signals that are critical to capture a given issue's full resolution steps.

\item \emph{L3: Inaccuracies in the Training Dataset:}
A common challenge in previous learning-based studies lies in the training phase and the preparation of training data. In ILR tasks, the only ground-truth data available are the actual issue-commit links. Consequently, researchers often generate artificial negative examples by assuming that any issue-commit pair not present in the ground truth is a false link. However, this assumption is flawed, as an issue may be resolved by multiple commits. In such cases, simply labeling each commit as true or false may not adequately capture this complexity. As a result, some commits labeled as negative in earlier studies may, in fact, represent true links~\cite{zhang2023ealink}. Such inaccuracies in the training dataset could impair performance in real-world scenarios.

\item \emph{L4: Impractical Pairwise Issue-Commit Approach:}
Most existing approaches focus on individual issue-commit pairs, and given a single issue-commit pair, they predict whether the commit is relevant to the issue~\cite{mazrae2021automated,ruan2019deeplink,lan2023btlink,lin2021traceability,zhang2023ealink,wang2025mplinker,deng2024promptlink}. This strategy becomes impractical for large repositories with tens of thousands of commits, as it requires generating and analyzing every possible issue-commit pair for a given issue before selecting the most relevant commit. Thus, the computational overhead and combinatorial explosion make these methods ill-suited for real-world, large-scale scenarios.

\end{enumerate}

\rev{To address these limitations, we redefine ILR as a dynamic heuristic search over the software repository's version-control graph that aggregates the full resolution logic of an issue. To achieve this, we present \ga{}, the first search-based ILR approach utilizing an autonomous LLM agent architecture. Given the issue URL and a link to the project repository, \ga{} grants the LLM on-demand access to various project data sources such as the commit history, issue threads, and codebase, via specialized function calls. \ga{} then explicitly instructs the agent to navigate the repository and selectively retrieve evidence to confirm or prune potential links. Through this heuristic search, the agent recovers the aggregate resolution logic contained within an entire commit chain without exceeding its token window.} The main contributions of our work are as follows:
\begin{itemize}


\item \rev{To overcome the limitations of one-to-one link recovery (i.e., addressing \emph{L1}), \ga{} employs an aggregate reasoning strategy that incorporates the entire commit chain during the decision-making process. Unlike prior models that examine commits in isolation, \ga{} iteratively explores the commit history, accumulating related changes until the \textit{aggregation} of all observed change logic and code diffs provides a complete resolution to the issue. 
This search-driven approach ensures that the agent does not prematurely terminate upon finding an intermediate fix; instead, it continues to traverse temporal and parental semantics until the cumulative code changes fully satisfy the issue requirements. 
Upon confirming that the gathered set of commits constitutes a total resolution, \ga{} returns the chronologically last commit within this set as the final resolution point. From this final commit, the entire resolution commit chain can be formally reconstructed by traversing the second-parent and chronological links~\cite{zou2019branch}. 
This shifting of focus from individual ``relevant'' commits to aggregate ``sufficient'' resolutions ensures higher precision and enables tasks that require the complete resolution history, such as accurately measuring issue resolution duration~\cite{panis2010successful}.}



\item \rev{To incorporate all available contextual data (i.e., addressing \emph{L2}), \ga{} replaces the static feature engineering used in prior work~\cite{huang2025back,zhang2023ealink,lin2021traceability,ruan2019deeplink,dong2022semisupervised} with a dynamic, lazy-access architecture. While existing methods are limited to pre-processed, high-level features, \ga{} leverages the interactive capabilities of LLM agents to perform on-demand data retrieval through specialized function calls.} \ga{} exposes APIs that the LLM can actively invoke to fetch only the specific information it deems necessary. \rev{This mechanism grants the model full visibility into data sources previously unavailable to ILR systems, such as codebase repositories and issue discussion threads, without exhausting the model's context window. By allowing the LLM to decide which ``slices'' of data are relevant at runtime, \ga{} can investigate semantic nuances that are lost in static embeddings.} Moreover, \ga{} selectively obtains feedback from the LLM after function calls to assess the relevance of the \rev{data}. If a \rev{data} is deemed unhelpful by the LLM, \ga{} discards it from the chat history to \rev{maintain the information density within the token window.}

\item Since \ga{} is built on a pre-trained, general-purpose LLM, it requires no task-specific training and remains robust against flaws in manually generated datasets (i.e., addresses \emph{L3}).


\item \rev{\ga{} reformulates ILR as a heuristic search problem, eliminating the need to exhaustively evaluate every commit in the repository (i.e., addressing \emph{L4}). By providing the LLM with on-demand access to the project's commit history, \ga{} allows the model to determine at runtime which chronological segments or specific commit attributes are most likely to contain the resolution logic. Instead of a linear scan, the agent selectively retrieves only the commits and metadata deemed essential for its reasoning process, effectively pruning irrelevant commits from the history. This targeted navigation allows \ga{} to converge on the final resolving commit and aggregate the contributing chain without the prohibitive cost of pairwise scoring.} 

\end{itemize}

We benchmarked \ga{} against five state-of-the-art ILR methods~\cite{huang2025back,ruan2019deeplink,lin2021traceability,dong2022semisupervised,zhang2023ealink}. Results show that \ga{} substantially outperforms existing methods in Hit@1, with improvements ranging from 41\% to 714\%. To assess the impact of contextual signals overlooked by prior work, we conducted an ablation study disabling different data sources. Results show that removing issue comments reduces \ga{}'s accuracy by 17\%, removing codebase access by 4\%, and removing both by 21\%. These drops confirm that issue comments and code inspection play a crucial role in \ga{}'s performance. 
\rev{As shown in our evaluations, \ga{} demonstrates \textit{consistent} results across diverse projects with \textit{minimal variance} by flexibly leveraging such sources at runtime, whereas prior methods exhibited noticeable performance fluctuations.}
To demonstrate generalizability and real-world applicability, we further tested \ga{} on \checkNum{120} randomly selected \github{} issues, resolved after the training cut-off date for \ga{}'s LLM. It successfully linked \checkNum{107} of them, highlighting its adaptability and real-world utility. Furthermore, our cost analysis shows that, for a given issue, \ga{} completes ILR in under 24 seconds at an average cost of \$0.01. Our evaluations focus on two popular issue-tracking systems: \jira{} and \github{}. However, \ga{}'s modular and flexible design makes it straightforward to extend to other platforms and incorporate new domain-specific data sources. Moreover, by leveraging the \tool{Tree-sitter} parser~\cite{tree-sitter}, \ga{} provides support for nearly all popular programming languages.  

The remainder of this paper is organized as follows. \secref{motiv} presents our motivating examples. \secref{method} describes the approach and implementation of \ga{}. \secref{eval} explains our evaluations and discusses the results. \secref{disc} discusses the implications of our findings and outlines future research directions. \secref{threats} addresses the potential threats to the validity of our findings. \secref{related} reviews related work. Finally, \secref{conclusion} concludes the paper.

\begin{figure*}[t]
    \centering
    \includegraphics[width=0.9\linewidth]{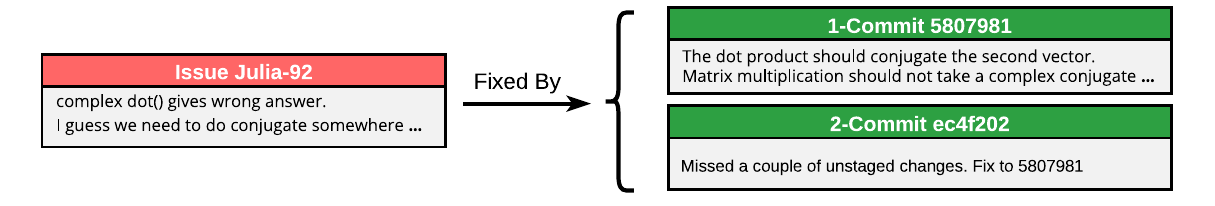}
    \vspace{-10pt}
    \caption{Motivating Example~1: Simplified commit timeline showing the initial commit \code{(5807981)}, which misses a minor change, followed by the final resolving commit \code{(ec4f202)}.}
    \label{fig:last_commit}
    \vspace{-10pt}
\end{figure*}

\begin{figure}[t]
    \centering

    \begin{subfigure}{\linewidth}
    \centering
    \includegraphics[width=0.9\linewidth]{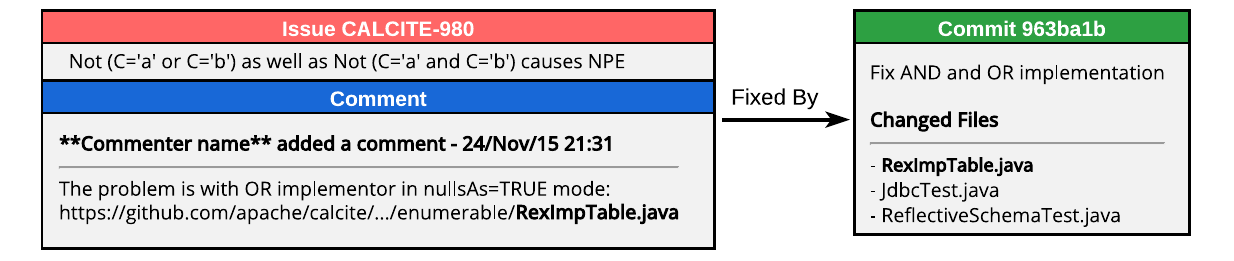}
        \caption{An example comment that urges a stronger solution and flags a file as the required modification site.}
        \label{fig:true_link}
    \end{subfigure}

    \begin{subfigure}{\linewidth}
        \centering
        \includegraphics[width=0.9\linewidth]{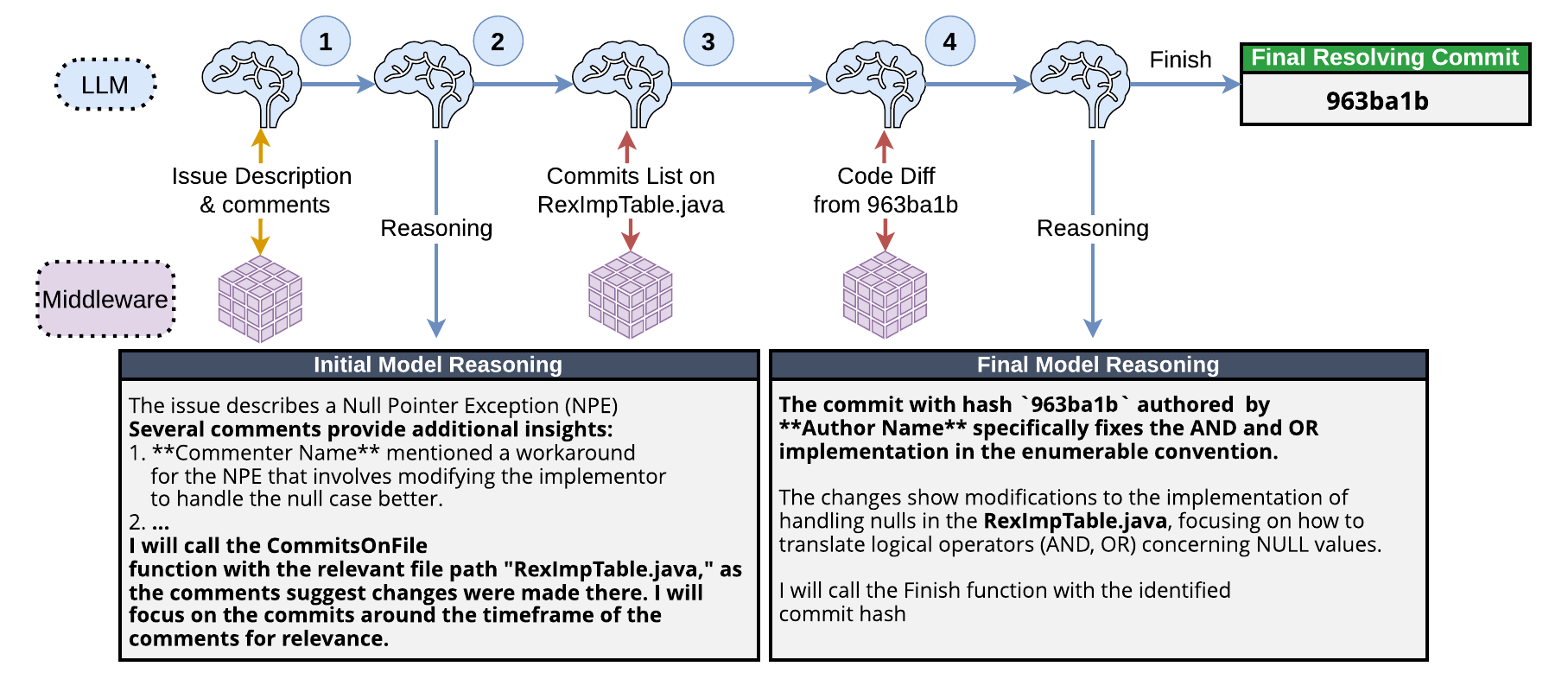}
        \caption{\ga{}'s complete workflow for retrieving the final resolving commit of issue CALCITE-980.}
        \label{fig:workflow}
    \end{subfigure}
    \vspace{-10pt}
    \caption{Motivating Example~2: Prior work's limited use of available context}
    \vspace{-10pt}

\end{figure}

\section{Motivating Examples}\label{sec:motiv}


\paragraph{\textbf{Example~1: The Context and Evaluation Gap.}}
\rev{State-of-the-art approaches typically treat ILR as an independent analysis of issue-commit pairs based on semantic similarity~\cite{huang2025back,zhang2023ealink,ruan2019deeplink,lin2021traceability,wang2025mplinker,deng2024promptlink}. In our examination of baseline failure cases, we observed that these methods often struggle with intermediate or final fixes whose commit messages are vague or brief (e.g., ``address code review'' or ``minor update'') and lack semantic overlap with the corresponding issue. Notably, these commits are semantically incomplete on their own. For instance, a commit may modify a function introduced in a previous relevant commit. Without including the previous commit in the ILR reasoning, the connection between the modification and the issue cannot be established. Our empirical observations suggest that this dependency on neglected relevant commits contributes to 18\% of the missed links in our top-performing baselines~\cite{zhang2023ealink,huang2025back}. Consequently, this isolated analysis fails to capture the cumulative nature of software fixes and the full resolution logic.} 

\figref{last_commit} shows an example of this in issue~\textit{Julia-92}\footnote{\url{https://github.com/JuliaLang/julia/issues/92}}, where an initial commit \code{(5807981)}\footnote{\url{https://github.com/JuliaLang/julia/commit/58079819d667f832abfc6fea8210252f161387d7}} explicitly claims to resolve the issue, but a subsequent commit \code{(ec4f202)}\footnote{\url{https://github.com/JuliaLang/julia/commit/ec4f20243c0807654eeeb8df343611d22c8ef404}} is required to provide the missing code changes required for true resolution. The baselines assign a high score to the initial commit based on its description, whereas the final commit receives a significantly lower score due to its nondescript message and the models' inability to process complex code diffs. Notably, because these models evaluate commits in isolation, they cannot recognize that the first commit is merely a partial fix and the true resolution is only achieved when the cumulative changes of the chain address the issue. In contrast, \ga{} iteratively examines the commit sequence until it confirms the aggregate changes provide a complete resolution, correctly identifying the final commit as the true link.

\paragraph{\textbf{Example~2: Limited Use of Available Context.}}
\rev{As discussed in Example 1, isolated issue-commit analysis often fails to acknowledge true links due to ambiguity or brevity in the issue text and commit messages. In our analysis of these failure cases, we observed that although the commit message and issue description may not align semantically, the issue discussions frequently include contextual cues about the changes being made. These hints often appear as specific references to file names, classes, or methods requiring updates, as well as explicit planning or explanations of the necessary changes and code snippets. Our empirical observations indicate that approximately 33\% of these specific missed links in top-performing baseline methods~\cite{zhang2023ealink,huang2025back} contained such signals in the issue comments, which baseline methods completely ignore.}

 A concrete example of this limitation is shown in \figref{true_link}, which illustrates the true link between issue \textit{CALCITE-980}\footnote{\url{https://issues.apache.org/jira/browse/CALCITE-980}} and its final resolving commit \textit{\code{963ba1b}}\footnote{\url{https://github.com/apache/calcite/commit/963ba1b1b3d2ab95989d8383e0a855c3ae5e24cb}}. The state-of-the-art ILR approaches, EasyLink~\cite{huang2025back} and EA-Link~\cite{zhang2023ealink}, both fail to recover this link because there is little semantic overlap between the issue report and the commit message. However, as shown in \figref{true_link}, the discussion thread contains a critical comment identifying the root cause in \texttt{RexImpTable.java} and suggesting a patch. 

\ga{} leverages these previously overlooked signals. As shown in \figref{workflow}, the agent first reads the entire discussion and infers that changes to \texttt{RexImpTable.java} are critical. It then searches the commit history for modifications to this file within the relevant timeframe. Among the resulting candidates, the agent inspects commit \code{963ba1b} in more detail, analyzing both its metadata and code diff. Based on this evidence, it concludes that commit \code{963ba1b} resolves the issue.


This example demonstrates two key advantages of \ga{}. First, it incorporates richer information sources (e.g., issue comments and codebase) that prior ILR methods ignore. Second, its workflow is not hard-coded: function calls are chosen adaptively by the agent in response to the context. This adaptivity enables more accurate recovery in cases where simple text similarity fails.




\section{The \ga{} Approach}\label{sec:method}

\begin{figure*}
    \centering
    \includegraphics[width=\linewidth]{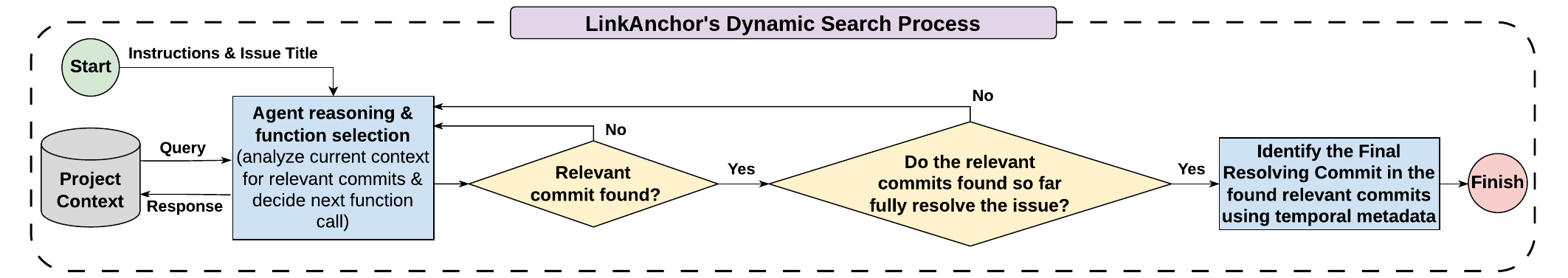}
    \vspace{-10pt}
    \caption{\rev{High-level workflow of \ga{}'s dynamic search}}
    \label{fig:search_process}
    \vspace{-10pt}
\end{figure*}

\begin{figure*}
    \centering
    \includegraphics[width=\linewidth]{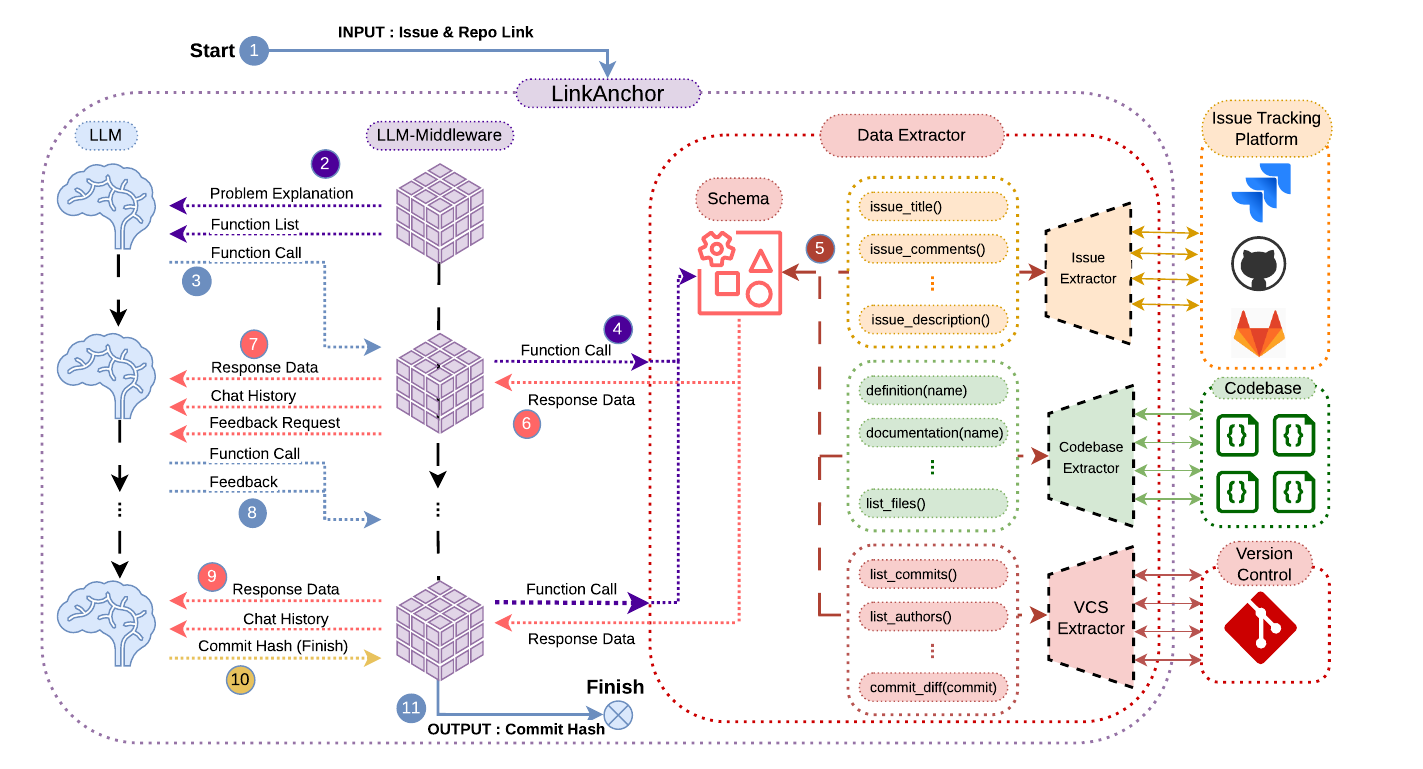}
    \vspace{-10pt}
    \caption{\ga{}'s overall architecture}
    \label{fig:ga-arch}
    \vspace{-10pt}
\end{figure*}

\rev{

\subsection{Dynamic Heuristic Search Paradigm}
\label{sec:goal-design}
Prior research shows that the relationship between issues and resolving commits is often one-to-many, meaning an issue usually requires multiple commits to fully resolve~\cite{zhang2023ealink,huang2025back}. However, existing ILR approaches typically evaluate commits in isolation and fail to capture the complete resolution logic when it is distributed across several commits (see~\secref{motiv}). To bridge this gap,~\ga{} redefines ILR as a dynamic heuristic search over the version control graph to incorporate the aggregate collection of commits whose combined changes fully satisfy the issue requirements. Unlike prior work, which ignores the temporal and parental relationships between commits, our search-based approach exploits the inherent associativity of software fixes.

\figref{search_process} illustrates~\ga{}'s dynamic search process. Instead of ingesting predetermined data points,~\ga{} interacts with the project as an explorable state space. Starting with the issue title, the agent follows an iterative reasoning process, evaluating the information gathered at each step to identify relevant commits and determine which specialized function to call next. This allows the agent to dynamically navigate the version control graph and formulate queries based on the evolving context. Rather than evaluating commits in isolation,~\ga{} accumulates context from relevant commits that may address different aspects of the issue.

This accumulation strategy is central to how~\ga{} identifies the final resolving commit in scenarios where multiple commits relate to the same issue. By employing a stopping condition based on aggregate reasoning, the agent continuously evaluates the cumulative changes of the relevant commits identified so far against the issue's requirements. This mechanism ensures that the search does not prematurely terminate upon discovering a partial fix. The search process continues until the agent determines that the aggregated relevant commits, when taken together, provide a complete resolution to the issue.

Once the agent confirms that the accumulated evidence constitutes a total resolution, it uses commit timestamps to distinguish the chronologically last commit among the identified relevant commits and returns it as the final issue resolution point. Notably, the complete resolving commit chain can then be formally reconstructed by traversing the second-parent links starting from this final resolving commit~\cite{zou2019branch}.}

\subsection{Architecture}
\label{sec:functions}
A key challenge faced by previous studies was condensing the vast amount of contextual information available to fit within the limited token window of the LLM. Critical data points, such as code snippets and issue comments, were often omitted due to their large size. With the emergence of LLM agents, this limitation can now be addressed by allowing the model to request exactly the information it needs on demand, rather than being provided the entire context upfront.

Building on this idea, \ga{} equips the LLM with a rich set of functions that grant access to various sources of relevant information, such as commit history, issue threads, discussions, and the codebase. Essentially, \ga{} frames the problem for the LLM and allows it to explore the given project by calling these functions to find the issue-commit link. When a function is invoked, \ga{} responds with the exact data needed.

As illustrated in \figref{ga-arch}, \ga{} comprises several modules. Given an issue URL and the corresponding Git repository, \ga{} initializes its data-extractor module, which manages all function calls for data retrieval. The data extractor supports three sources: 
(i) Issue data, such as the title, description, and discussion threads;
(ii) Queries over the VCS and commit history, such as retrieving lists of commits filtered by specific attributes, and
(iii) Advanced code navigation, such as retrieving the documentation and definitions for methods or classes in the codebase. 

Each source has an extractor, which registers its supported APIs (functions) in a shared schema. This schema is later used to route each call issued by the LLM to the appropriate extractor.
Because adaptability across diverse environments is a core requirement, \ga{} should be designed for modularity and pluggability. This architecture allows new data-extractor modules to be plugged into the system simply by extending the schema at runtime.

The LLM-Middleware handles communication with the LLM via API calls. It first prompts the LLM with the problem explanation and the available functions (from the data-extractor's schema). Retrieval then proceeds as an iterative dialogue between the LLM and the middleware.

LLM analyzes the provided information and takes one of the following actions: (i) invokes \code{Finish} to mark a commit as the resolution;
(ii) requests additional contextual data via the available functions;
(iii) requests a different batch of commits; and
(iv) invokes \code{GiveUp} to terminate due to insufficient data.

These iterations continue until the LLM either finds the correct commit or terminates due to insufficient contextual data (\code{GiveUp}).
To ensure termination, a hard limit is set on the number of iterations, and this number is communicated to the LLM in the initial prompt.

In total, \ga{} offers 20 functions, matching the currently recommended maximum for models like ChatGPT-4o~\cite{openaiFunctionCalling2024}. All functions are implemented deterministically, so every response from \ga{} can be verified against the true project data.

As shown in \tabref{api-functions}, \ga{} provides four categories of functions for the LLM: Git, Issue, Codebase, and Control. The first three grant the LLM access to information about the Git history, issues, and the codebase, respectively. Control functions manage the flow of communication with the LLM.

\begin{table}[t]
  \centering
  \scriptsize
  \begin{threeparttable}    
    \caption{API Function Definitions for \ga{}}
    \label{tab:api-functions}
    \vspace{-0.4cm}
    \begin{tabular}{llll}
      \toprule
      \textbf{Function} & \textbf{Inputs} & \textbf{Output} & \textbf{Description} \\
      \midrule
      \multicolumn{4}{l}{\textbf{Git Functions}} \\
      \midrule
      \mbox{list\_commit} & \mbox{pagination} & \mbox{List[CommitMeta]} & Paginated commits \\
      \mbox{list\_authors} & — & \mbox{List[Author]} & List of authors \\
      \mbox{authors\_commits} & \mbox{Author, pagination} & \mbox{List[CommitMeta]} & Commits by the specified author \\
      \mbox{list\_files} & \mbox{pattern} & \mbox{List[Path]} & Files in history matching pattern \\
      \mbox{commits\_on\_file} & \mbox{file\_name, pagination} & \mbox{List[CommitMeta]} & Commits that staged specified file \\
      \mbox{commit\_diff} & \mbox{Commit\_hash} & \mbox{CommitDiff} & Diff of the specified commit \\
      \mbox{commit\_metadata} & \mbox{commit\_hash} & \mbox{CommitMeta} & Metadata: author, msg, timestamps \\
      \midrule
      \multicolumn{4}{l}{\textbf{Issue Functions}} \\
      \midrule
      \mbox{issue\_title} & — & \mbox{String} & Title of the current issue \\
      \mbox{issue\_description} & — & \mbox{String} & Description of the issue \\
      \mbox{issue\_created\_at} & — & \mbox{DateTime} & Issue creation timestamp \\
      \mbox{issue\_closed\_at} & — & \mbox{DateTime} & Issue resolution timestamp \\
      \mbox{issue\_author} & — & \mbox{Author} & Username of the issue's author \\
      \mbox{issue\_comments} & \mbox{pagination} & \mbox{List[Comment]} & Comments on the issue \\
      \mbox{issue\_participants} & — & \mbox{List[Author]} & Users participated in issue thread \\
      \midrule
      \multicolumn{4}{l}{\textbf{Codebase Functions}} \\
      \midrule
      \mbox{fetch\_definition} & \mbox{commit, path, name} & \mbox{String} & Source code of a function/class \\
      \mbox{fetch\_document} & \mbox{commit, path, name} & \mbox{String} & Docstring of a function/class \\
      \mbox{fetch\_lines\_in\_file} & \mbox{commit, file, start, end} & \mbox{String} & Range of lines from a file \\
      \midrule
      \multicolumn{4}{l}{\textbf{Control Functions}} \\
      \midrule
      \mbox{finish} & \mbox{commit\_hash} & — & Mark resolving commit and terminate \\
      \mbox{Give\_up} & — & — & Give up and terminate the process \\
      \mbox{feedback} & \mbox{id, Discard/Store} & — & Provide feedback on a function call \\
      \bottomrule
    \end{tabular}
    \begin{tablenotes}
      \small
      \scriptsize
      \item \textit{Note:} The \mbox{commit} input in the codebase functions is used to check out the codebase at a specific commit\\ before retrieving the requested data (e.g., function code, docstring, class attributes).
    \end{tablenotes}
  \end{threeparttable}
    \vspace{-13pt}
\end{table}

\paragraph{\textbf{Git Functions.}} 
Enable the LLM to execute rich, complex queries on the commit history of the repository. We divide them into two categories: 
(i) Batch-retrieval functions that return a set of commits filtered by attributes such as author, file changes, or time period. These functions always return the commit metadata information, including the commit message, author, commit hash, and commit timestamp.
(ii) Data-retrieval functions, which supply parameters for those batch queries. For example, the LLM can call \code{list\_authors} to obtain all commit authors, then pass a selected author to \code{commits\_of\_author} to fetch that author's commits.

When faced with a large repository, the full context of the repository can overwhelm the LLM and degrade decision accuracy, since the full commit history often consists of tens of thousands of commits across different branches.
To address this issue, the Git-Extractor module provides a much simpler view of the Git repository to the LLM. In particular, the Git-Extractor module provides two key abstractions:

\begin{enumerate}
\item Unified branch view: The Git-Extractor module provides the illusion of having a single master branch to the LLM by merging all branches into one in order to reduce complexity for the LLM. This way, the LLM sees one coherent sequence of commits. 
This is especially important since it is common practice to maintain separate branches for each version of the artifact, and this separation does not provide any useful data for the LLM and only causes confusion.

\item ``Safe-lifespan'' filtering: The Git-Extractor module limits all queries to the safe lifespan of the issue, defined as one week before issue creation through one week after issue resolution. Prior studies show that most resolving commits fall within this time window~\cite{ruan2019deeplink,sun2017frlink,zhang2023ealink}. This way, when calling functions like \code{commits\_of\_author} or \code{list\_files}, only commits that are within the safe life span of the issue are considered for responding to the query. 
Limiting the search space in this way both reduces token consumption and speeds up the link‐retrieval process. Nevertheless, the LLM can still examine a commit outside of the default time period by calling the \code{list\_commits} function. However, the default is set to the safe lifespan.
\end{enumerate}

\paragraph{\textbf{Codebase Functions.}} Allow the LLM to retrieve the documentation and/or implementation of specific methods, classes, or data structures. They also support fetching arbitrary lines of code from a given file. This enables the LLM to explore the codebase and identify potential locations where a committed change may address and resolve the issue. Since the state of the codebase can change with each commit, all codebase functions take a \code{commit\_hash} as input. Together, these functions provide the LLM with a precise view of the code at a specific point in time.

In order to support various languages, \ga{} leverages Tree-sitter~\cite{tree-sitter} as an abstract syntax tree (AST) parser. As a result, \ga{} needs to define only two new Tree-sitter queries per language: one for retrieving data structures and another for retrieving functions.
Because Tree-sitter already supports most of the popular languages, \ga{}'s support can be extended to a new language by simply writing the two mentioned Tree-sitter queries.

\paragraph{\textbf{Issue Functions.}} Through these functions, the LLM accesses issue details such as the title, description, and discussion threads. To avoid exhausting the token window, the comment retrieval function supports pagination.

\paragraph{\textbf{Control Functions.}} Are useful for controlling the flow of the process. Currently, there are three functions defined for the LLM to use. \code{Finish} signals successful completion by returning the commit hash of the identified commit. \code{GiveUp} indicates that the LLM could not pinpoint the target commit. Whenever a function call returns more than a predefined number of bytes of data, a request for Feedback is issued from \ga{}. In these cases, the LLM should invoke \code{Feedback} with either \code{Discard}, to signal that the data should be replaced with a brief ``omitted'' notice, or \code{Preserve}, to keep the original output. By discarding irrelevant or excessively large data, we prevent needless token consumption and keep the iterative process moving forward.

To support performance tuning and future enhancements, \ga{} includes a metrics module that automatically records every function call made by the LLM. For each issue-commit recovery session, it captures two types of data: \emph{Function-call counts}, which represent the total number of times each function is invoked, and \emph{Call sequences}, which are the ordered lists of functions the LLM requests.
We use these metrics in our evaluation (\secref{eval}).


\subsection{Prompting}
\label{sec:prompting}
\rev{The initial prompt serves as the foundational logic for~\ga{}, governing how the LLM approaches the ILR task and establishing the rules for its operation in practical settings. To operationalize the design described in~\secref{goal-design}, the prompt explicitly frames ILR as a dynamic search process. We achieve this by giving instructions that require the agent to iteratively accumulate context from the commit history and evaluate whether the aggregated changes are sufficient to fully resolve the issue.

We explicitly inform the agent that developers often distribute a single fix across a sequence of commits; therefore, the agent should not stop at the first relevant commit it encounters. Instead, the model is directed to continue its search until the \textit{cumulative} changes across all identified relevant commits provide a complete resolution to the issue. Once this state of total resolution is confirmed, the agent is tasked with reporting the chronologically last commit in that chain as the final resolving link. To support this, the prompt instructs the LLM to prioritize the analysis of actual code diffs over potentially vague or misleading commit messages. The specific instructions included in the agent's initial system prompt that enforce these behaviors are shown below:
}
\definecolor{mytitlecolor}{HTML}{435066}
\newtcolorbox{promptbox}[1]{
colback=gray!10,        
colframe=black,         
coltitle=white,         
colbacktitle=mytitlecolor,     
fonttitle=\small\bfseries,    
center title,           
title=#1,               
boxrule=0.8pt,
arc=4pt,                
left=6pt,
right=6pt,
top=6pt,
bottom=6pt
}

    \begin{promptbox}{System Prompt Guidelines for Multi-Commit Fixes \& Code Diff Prioritization}
    \rev{\ttfamily\scriptsize
    [...]\newline
    When resolving an issue, a developer might split the fix across multiple commits. In your search, you might come across commits that are partial fixes. You should continue your search until the cumulative changes from all identified relevant commits fully resolve the issue. Finally, use temporal metadata to return the last commit hash in the sequence as the resolving commit.
    \newline[...]\newline
    Commit messages may not fully reflect the changes. Do not solely rely on the message to determine if a commit resolves the issue. Instead, analyze the actual code changes (diffs) to make an informed decision.}
    \end{promptbox}

To ensure termination, the LLM is informed in the initial prompt that it has a limited number of function calls and must generate a final result before reaching this limit. 
All of our prompts are available in our replication package~\cite{Git-Anchor}.

\section{Evaluation}\label{sec:eval}
In this section, we evaluate how effective \ga{} is in the issue-to-commit link recovery task. We aim to answer the following research questions (RQs):
\begin{itemize}
\item \textbf{RQ1}: How well does \ga{} perform vs. baselines?
\item \textbf{RQ2}: What is the contribution of each data source to issue-commit link recovery?
\item \textbf{RQ3}: How well does \ga{} perform on new, unseen, real-world data?
\item \textbf{RQ4}: What are the costs of the approach?
\end{itemize}

\subsection{\textbf{RQ1: How well does \ga{} perform vs. baselines?}}\label{sec:EvalBaselines}
We evaluate the performance of \ga{} relative to state-of-the-art baselines in this section.

\subsubsection{Experimental Setup.}
\label{sec:eval_metrics}

\paragraph{\textbf{Dataset.}}\label{sec:dataset}
We evaluate~\ga{} on the dataset introduced in~\cite{zhang2023ealink}, which includes over 56k unique issues and all of their true links from six Apache projects. All these projects use Jira for issue tracking and cover a total of 27 unique git repositories, enabling a thorough evaluation of \ga{} on diverse project environments. \rev{Furthermore, approximately 46\% of the issues in this dataset have multiple resolving commits, ensuring that multi-commit fixes are well-represented in our evaluation.}

\paragraph{\textbf{Evaluation Metrics.}}
State-of-the-art approaches have typically framed ILR as a recommendation problem~\cite{huang2025back,lin2021traceability,ruan2019deeplink,dong2022semisupervised}, where the system produces a top-$k$ ranked list of candidate commits for each issue, and evaluation is based on whether any of the true links in the dataset appear within the top-$k$ results. This is typically measured by the Hit@k metric, which captures the likelihood that at least one related commit appears within the retrieved set.
\begin{equation}
\mathrm{Hit@}k \;=\;
\frac{1}{\lvert Q\rvert}
\sum_{i=1}^{\lvert Q\rvert}
\mathbb{I}\bigl(\mathrm{Rank}_i \le k\bigr)
\label{eq:hitk}
\end{equation}

In Equation~\ref{eq:hitk}, $Q$ is the query set, with $|Q|$ denoting its size. Furthermore, $\mathbb{I}$(·) returns 1 if any true link (i.e. any relevant commit) for query $i$ is within the top-$k$, and 0 otherwise.

\ga{} formulates ILR as the task of retrieving the last chronological resolving commit for each issue. Since the last resolving commit is inherently included among the set of ``true links'' in the dataset, this formulation allows us to adopt state-of-the-art ILR recommendation methods as baselines and to follow their established evaluation protocols. In doing so, we ensure meaningful and direct comparison.~\rev{Notably, approaches such as MPLinker~\cite{wang2025mplinker} and PromptLink~\cite{deng2024promptlink} frame ILR as a binary classification task. Given a single isolated issue-commit pair, these methods predict a binary label (True-Link or False-Link). However, because this formulation inherently lacks a relative scoring mechanism to rank one valid commit over another, it is incompatible with the ranking-based evaluation protocol used in this study. Consequently, these methods are excluded from our baselines.}

The most appropriate value for k for evaluating \ga{} against the baselines is one. Hit@1 metric evaluates the probability that the most related (top-ranked) commit is a true link. We compute \ga{}'s Hit@1 as the fraction of issues for which the retrieved commit hash matches the final resolving commit. Notably, comparing \ga{}'s Hit@1 with the Hit@1 reported by our baselines puts \ga{} at a significant disadvantage: when multiple commits resolve an issue, baselines receive credit for retrieving any commit in the resolving chain, whereas \ga{} is only counted as correct if it identifies the final commit. As a result, a commit that is relevant but non-final is scored as a hit for the baselines but a miss for \ga{}.
Beyond this primary comparison, we also conduct a secondary comparison of \ga{}'s Hit@1 against the baselines' Hit@10 results. This imposes an even stricter standard, as Hit@10 is by definition always greater than or equal to Hit@1, since allowing more candidates in the retrieved set increases the likelihood of including a correct commit. Nonetheless, we adopt these criteria to demonstrate that \ga{} can remain competitive even under evaluation conditions that inherently favor the baselines. 

\paragraph{\textbf{Model Configuration.}}
We used OpenAI's ChatGPT-4o-mini~\cite{OpenAIGPT4o2025} model as the backend LLM, which offers fast performance while maintaining reasonable quality. \change{For evaluation purposes, we filtered out all issue keys from the data exposed to the LLM.} \checkNum{The context window size for the LLM is 128k tokens~\cite{docsbot_gpt4o_token_window_limit}.} We limited our agent to a maximum of 20 iterations when communicating with the LLM and restricted it to a total of 200k tokens per issue. The size limit for the Feedback function is set to 40kB, which roughly translates to 10k tokens per iteration~\cite{OpenAITokens2025}. These settings enable us to have control over the overall cost and time of ILR for each issue. But this also means that for some issues, \ga{} does not provide any resolving commits.
Moreover, LLMs can be unpredictable, and their responses are not always reproducible. To mitigate the potential noise introduced by this, we repeated the experiment three times and report the average results.

\paragraph{\textbf{Baselines.}}
Below, we provide a brief description of these baselines:

\begin{itemize}
    \item \emph{EasyLink}~\cite{huang2025back} retrieves the top-$10$ candidate commits using a FAISS~\cite{douze2024faiss} vector database and then applies an LLM to rerank them by relevance.
    
    \item \emph{EALink}~\cite{zhang2023ealink} distills knowledge from CodeBERT~\cite{feng2020codebert} into a smaller model, fine-tuned with multi-task contrastive learning.
    
    \item \emph{T-BERT}~\cite{lin2021traceability} is first pre-trained on the large-scale CodeSearchNet~\cite{husain2019codesearchnet} dataset for the code search task~\cite{liu2022opportunities}, and then fine-tuned for the link recovery task.
    
    \item \emph{DeepLink}~\cite{ruan2019deeplink} is a DL based ILR method that uses RNNs to learn the semantic representations of issues and commits.
    
    \item \emph{VSM}~\cite{dong2022semisupervised} is an IR method that expresses textual data as word vectors. The relevance between an issue and a commit is calculated based on the similarity between their corresponding word vectors.
\end{itemize}

\begin{table}[t]
  \centering
  \caption{Hit@1 Scores Across Six Projects (``Improvement'' is \ga{}'s mean gain over each baseline)}
  \label{tab:hit1_single_reordered}
  \vspace{-0.4cm}  
  \scriptsize
  \setlength{\tabcolsep}{4pt}
  \begin{tabular}{lccccccc}
    \toprule
    \textbf{Project}
      & \textbf{\ga{}}
      & \textbf{EasyLink}
      & \textbf{EALink}
      & \textbf{T-Bert}
      & \textbf{DeepLink}
      & \textbf{VSM} \\
    \midrule
    Ambari     & \textbf{0.9546} & 0.7812 & 0.9490   & 0.5524   & 0.0625   & 0.4209   \\
    Calcite    & \textbf{0.8874} & 0.8654 & 0.6555   & 0.5540   & 0.0926   & 0.2770   \\
    Groovy     & \textbf{0.8424} & 0.4590 & 0.6650   & 0.4655   & 0.0062   & 0.4983   \\
    Ignite     & \textbf{0.8221} & 0.4977 & 0.3950   & 0.4304   & 0.0584   & 0.1012   \\
    Isis       & \textbf{0.8827} & 0.3808 & 0.2423   & 0.2394   & 0.3291   & 0.1191   \\
    Netbeans   & \textbf{0.7901} & 0.6357 & 0.3273   & 0.2794   & 0.0870   & 0.0125   \\
    \midrule
    Mean       & \textbf{0.8632} & 0.6118 & 0.5390   & 0.4202   & 0.1060   & 0.2382   \\
    \textbf{Improvement}
               & \textbf{—}
               & \textbf{41.09\%}
               & \textbf{+60.15\%}
               & \textbf{+105.43\%}
               & \textbf{+714.61\%}
               & \textbf{+262.44\%} \\
    \rev{\textbf{Standard Deviation ($\sigma$)}}
               & \rev{\textbf{0.0532}}
               & \rev{0.1776}
               & \rev{0.2394}
               & \rev{0.1167}
               & \rev{0.1021}
               & \rev{0.1746} \\
    \bottomrule
  \end{tabular}
\end{table}

\begin{table}[t]
  \centering
  \caption{Comparison of Baseline Hit@10 vs.\ \ga{}'s Hit@1 Across Six Projects}
  \label{tab:hit1_vs_hit10}
  \vspace{-0.4cm}  
  \scriptsize
  \setlength{\tabcolsep}{4pt}
  \begin{tabular}{lcccccc}
    \toprule
    \textbf{Project}
      & \textbf{\ga{}}
      & \textbf{EasyLink}
      & \textbf{EALink}
      & \textbf{T-Bert}
      & \textbf{DeepLink}
      & \textbf{VSM}
      \\
    \midrule
    Ambari     & 0.9546 & 0.9310 & \textbf{0.9800} & 0.7748 & 0.2167 & 0.6798  \\
    Calcite    & 0.8874 & \textbf{0.9474} & 0.8566 & 0.8357 & 0.1739 & 0.7512  \\
    Groovy     & \textbf{0.8424} & 0.7570 & 0.8310 & 0.7175 & 0.1522 & 0.7763  \\
    ignite     & \textbf{0.8221} & 0.6770 & 0.7210 & 0.7087 & 0.1700 & 0.5806  \\
    Isis       & \textbf{0.8827} & 0.6334 & 0.5184 & 0.4304 & 0.4960 & 0.4705  \\
    Netbeans   & \textbf{0.7901} & 0.7887 & 0.6727 & 0.5074 & 0.1925 & 0.3750  \\
    \midrule
    Mean       & \textbf{0.8632} & 0.79075 & 0.7633 & 0.6624 & 0.2336 & 0.6056  \\

    \bottomrule
    \vspace{-17pt}
  \end{tabular}
\end{table}



\subsubsection{Experimental Results.}
\tabref{hit1_single_reordered} reports each method's Hit@1 performance across all six projects. 
\ga{} achieves the highest Hit@1 on every project, outperforming EasyLink, by up to 231\% on the Isis project. 
To further highlight \ga{}'s effectiveness, \tabref{hit1_vs_hit10} compares \ga{}'s Hit@1 results with the other models' Hit@10 results. 
All competing methods show substantial score gains when moving from Hit@1 to Hit@10, since a larger K typically yields a higher Hit@K by increasing the chance of including the resolving commit in the top-K list.
\ga{}'s single prediction still outperforms every other method's Hit@10, except on the Ambari project, where EA-Link's Hit@10 exceeds \ga{}'s Hit@1 by only 0.025 and Calcite where EasyLink outperforms \ga{} by 0.06. 
It is notable that \ga{} achieves these results without the cost and overhead of training, dataset feature-engineering, or extensive hardware requirements, unlike the other methods. 
Furthermore, \ga{} shows a high, steady, and robust Hit@1 score across all projects (between 0.79 and 0.95 \rev{with a standard deviation of 0.05}), whereas other methods' Hit@1 scores fluctuate based on the project context. For example, EasyLink's Hit@1 shows fluctuation over a wider range, between 0.38 (the Isis project) and 0.86 (the Calcite project), \rev{with a standard deviation of 0.17}.


\begin{tcolorbox}
\ga{} consistently outperforms all baselines in Hit@1. In particular, \ga{} achieved the highest Hit@1 score across every evaluated project, with average improvements ranging from 41\% to 714\% over existing methods. Moreover, \ga{} surpassed the baselines' Hit@10 performance in nearly all cases, highlighting its ability to deliver superior accuracy even on metrics that substantially favor the baselines.
\end{tcolorbox}

The performance variability of the baseline methods 
can be attributed to their reliance on a fixed set of features, such as issue summaries, descriptions, and commit messages. This could potentially be problematic since the availability, quality, and effectiveness of these features vary widely across different repositories. For example, issue descriptions and commit messages can differ in length, detail, and writing style from one project to another~\cite{zhang2023ealink}. This dependence on static features prevents these approaches from generalizing well and leads to the performance fluctuations we observe in their results. \ga{}, on the other hand, requires no training or feature engineering. Instead, it can choose at runtime which data to inspect from all available sources and discard irrelevant information so as not to exceed its token span. This prevents \ga{} from relying too heavily on certain features. It can flexibly explore different data sources and focus on each one individually, without the noise introduced by loading everything into a single prompt. To further illustrate this, we propose the second research question.

\subsection{\textbf{RQ2: What is the contribution of each data source to issue-commit link recovery?}}
\label{sec:rq4}

\subsubsection{Experimental Setup.}
\ga{}'s modular design allows fine-grained control over the data sources it can access. To study the impact of each source, we design three ablation experiments using the same dataset as before. The only difference is that in each experiment, certain data sources are disabled. Specifically, in the first experiment, issue comments are removed; in the second, the codebase is removed; and in the third, both the codebase and issue comments are removed. The third setup resembles prior works more closely, since most of them had no access to either code files or issue comments.

\paragraph{\textbf{Dataset and Evaluation Metrics.}}
We conduct this ablation study
using the same evaluation metrics and dataset described in RQ1. We select this dataset because its large size and diverse project environments provide a strong basis for examining the individual impact of each data source on our approach.

\begin{figure*}[t]
    \centering
    \includegraphics[width=0.55\linewidth]{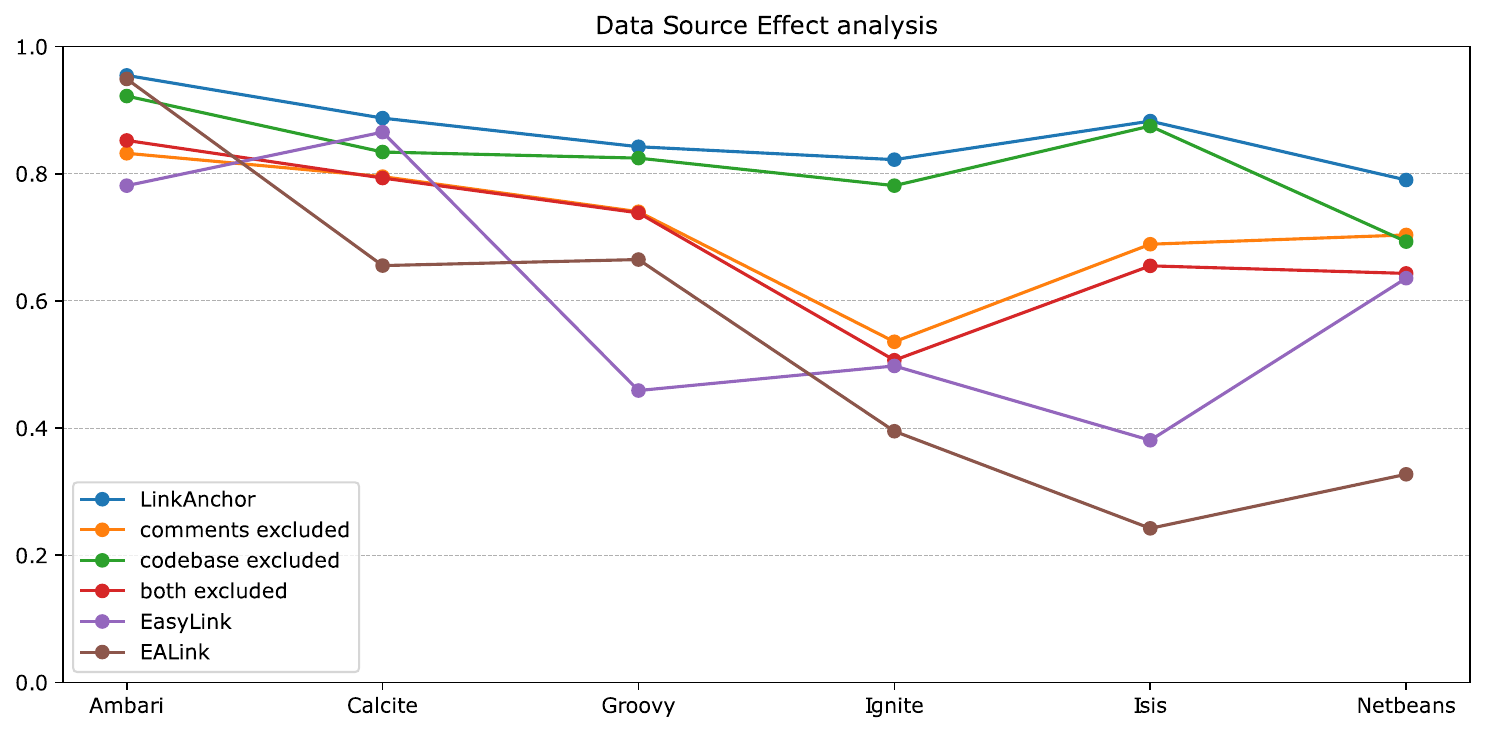}
    \vspace{-5pt}
    \caption{\ga{}'s performance in the absence of various data sources}
    \label{fig:ablation}
    \vspace{-10pt}
\end{figure*}

\subsubsection{Experimental Results.}
\figref{ablation} reports the results. On average, \ga{}'s accuracy drops by 21\% when both comments and code files are removed. Removing only comments leads to a 17\% drop, while removing only the codebase results in a 4\% drop. As shown in \figref{ablation}, the impact of these sources varies across projects, reflecting differences in data quality. For instance, \ga{}'s accuracy decreases by 34\% on the Ignite project when comments are removed, and up to 7\% on the NetBeans project when code files are removed. These results demonstrate that the observed accuracy gains are a direct result of \ga{}'s ability to flexibly choose and combine data sources at runtime.

\begin{figure*}[t]
  \centering
  \begin{subfigure}{0.32\textwidth}
    \includegraphics[width=\linewidth]{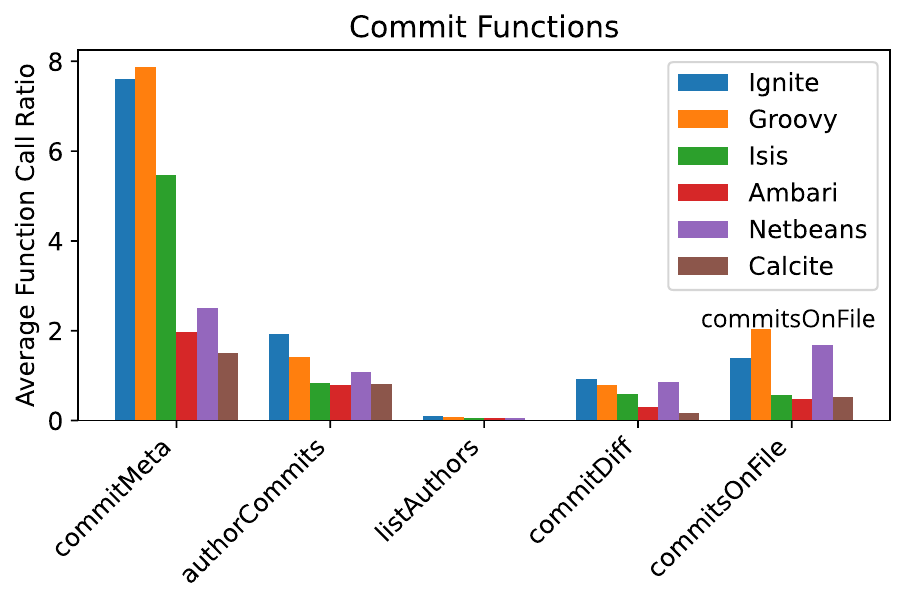}
    \vspace{-5pt}
    \caption{Git‐related function calls}
    \label{fig:metrics-git}
  \end{subfigure}
  \begin{subfigure}{0.32\textwidth}
    \includegraphics[width=\linewidth]{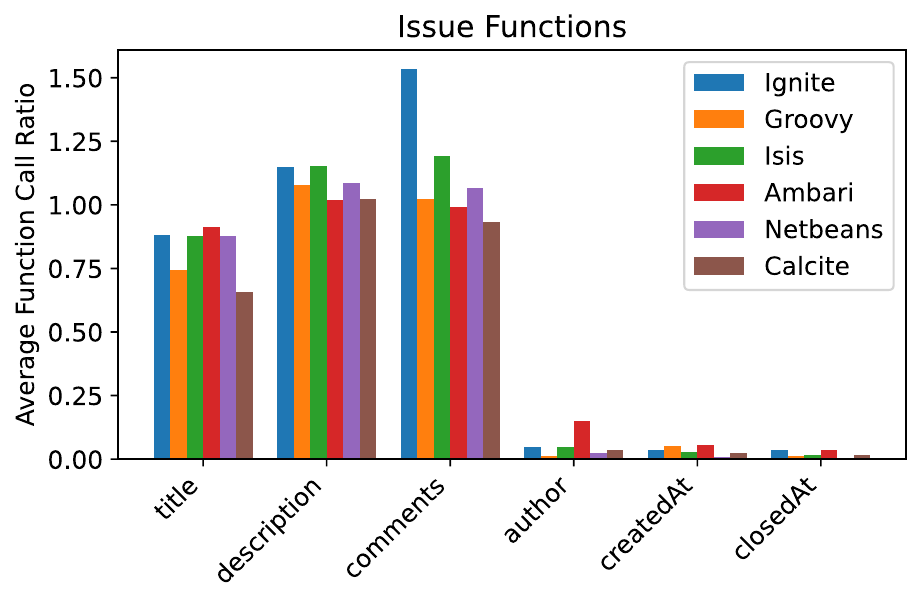}
    \vspace{-5pt}
    \caption{Issue‐related function calls}
    \label{fig:metrics-issue}
  \end{subfigure}
  \begin{subfigure}{0.32\textwidth}
    \includegraphics[width=\linewidth]{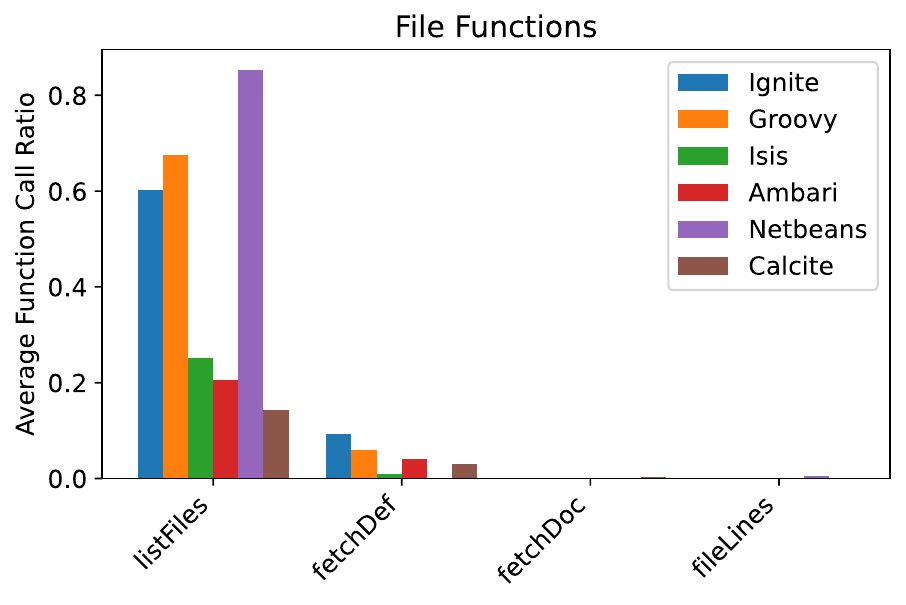}
    \vspace{-5pt}
    \caption{Codebase function calls}
    \label{fig:metrics-codebase}
  \end{subfigure}
  \vspace{-5pt}
  \caption{Average function‐call ratio per project \change{(Note that function names have been condensed for clarity, and the y-axis scales vary across subfigures, so they should not be directly compared)}.}
  \vspace{-10pt}
  \label{fig:all-metrics}
\end{figure*}

\figref{all-metrics} further supports this observation by presenting the average number of calls per function per issue in the original experiment (with all data sources enabled). Two key insights emerge:

\begin{enumerate}
\item \emph{Project‐Specific Data Importance}. Certain functions are used more heavily in specific projects, such as \texttt{issue\_comments} in Ignite and \texttt{list\_files} in NetBeans. This aligns with the performance drops observed earlier, indicating that these functions provided particularly valuable information in those contexts. This behavior confirms that \ga{}'s flexibility allows it to adapt to the unique characteristics of each project. Notably, previous methods struggled with these projects—for example, EasyLink and EALink achieved only \checkNum{0.38} and \checkNum{0.32} Hit@1 scores on Ignite, which further highlights the advantage of \ga{}'s dynamic data selection.
\item \emph{Effectiveness of Previously Untapped Data Sources}. Functions accessing data unavailable to earlier approaches, particularly \texttt{issue\_comments} and the suite of codebase‐inspection functions, show substantial usage (averaging 1.12 calls per issue for \texttt{issue\_comments}, 0.45 for \texttt{list\_files}, and 0.98 combined for all codebase functions). These numbers highlight the value of richer information channels for the ILR task and explain why \ga{} achieves consistently strong, stable performance across diverse repositories.

\end{enumerate}

\begin{tcolorbox}
Ablation experiments show that removing issue comments leads to a 17\% drop in accuracy, removing codebase access by 4\%, and removing both by 21\%, demonstrating the crucial role of these data sources in \ga{}'s performance on ILR tasks.
\end{tcolorbox}

\subsection{\textbf{RQ3: How well does \ga{} perform on new, unseen, real-world data?}}
\label{sec:rq3}

\begin{table}[t]
\centering
\begin{minipage}{0.48\textwidth}
  \centering
  \scriptsize
  \caption{\ga{} Hit@1 Performance}
  \label{tab:gitanchor_hit1}
  \vspace{-0.4cm}  
  \setlength{\tabcolsep}{8pt}
  \begin{tabular}{llrr}
    \toprule
    \textbf{Language} & \textbf{Project}                & \textbf{Stars} & \textbf{Hit@1} \\
    \midrule
    \multirow{4}{*}{Python}
      & huggingface/transformers          &  145k & 0.95  \\ 
      & pallets/flask                     & 69.6k & 0.95 \\ 
      & pytorch/pytorch                   & 24.2k& 0.90 \\ 
      & \textbf{Python Overall (mean)}  &  & \textbf{0.93} \\
    \midrule
    \multirow{4}{*}{Go}
      & golang/go                         & 128k & 0.85 \\ 
      & kubernetes/kubernetes             & 115k & 0.90 \\ 
      & moby/moby                         & 69.8k & 0.80 \\ 
      & \textbf{Go Overall (mean)}      &  & \textbf{0.85} \\
    \midrule
      & \textbf{All Projects (mean)}    &  & \textbf{0.89} \\
    \bottomrule
  \end{tabular}
\vspace{-0.4cm}
\end{minipage}%
\hfill
\begin{minipage}{0.48\textwidth}
\centering
\scriptsize
\vspace{-0.4cm}
\caption{\ga{} Cost Analysis Per Project}
\vspace{-0.4cm}
  \label{tab:costs}
  \begin{tabular}{lrrr}
    \toprule
    \textbf{Project} & \textbf{Time (s)} & \textbf{Token Usage} & \textbf{Cost (USD)} \\
    \midrule
    Isis     & 29.135463 & 123,042.06 & 0.0108 \\
    Calcite  &  9.983499 & 42,074.27 & 0.0037 \\
    Ambari   & 12.600633 & 68,782.48 & 0.0060 \\
    Groovy   & 41.536681 & 198,117.73 & 0.0174 \\
    Netbeans & 20.931454 & 150,424.12 & 0.0132 \\
    Ignite   & 27.403428 & 109,333.76 & 0.0096 \\
    \midrule
    \textbf{Average} & 23.598526 & 115,295.73 & 0.0101 \\
    \bottomrule
  \end{tabular}
\end{minipage}
\end{table}

\subsubsection{Experimental Setup.}
This research question aims to assess \ga{}'s performance against unseen data from real-world projects and analyze its generalizability. The training cut-off date for \ga{}'s LLM (Chat-GPT4o-mini) is October 2023. Therefore, to eliminate any risk of data leakage, we prepared a dataset of 120 randomly selected issues from six different repositories (20 issues from each), where the issue resolution and commit times for all issues occurred after October 2023. These issues are drawn from prominent open-source software projects hosted on GitHub. Three of the selected projects use Python, and the other three use Go as their primary programming language. \tabref{gitanchor_hit1} presents the list of selected projects and their corresponding GitHub stars.

\paragraph{\textbf{Evaluation Metrics.}}
To evaluate \ga{}'s generalizability and performance on new and unseen data, we adopt the same metric used in RQ1. We treat each issue in our test set as a single trial: if \ga{}'s predicted commit matches the last resolving commit in our dataset, we mark the trial as successful. We then calculate the overall success rate by dividing the number of successful trials by the total number of issues.

\subsubsection{Experimental Results.}
\tabref{gitanchor_hit1} presents \ga{}'s results on the constructed dataset. \ga{} achieved a mean Hit@1 score of 0.89 across all projects. These results demonstrate \ga{}'s strong generalizability to new, unseen data. Similar to the results achieved in RQ1, \ga{} has consistently high and robust accuracy across every project in the dataset, varying between a range of 0.80 and 0.95. Together, the results of RQ1 and RQ3 show that \ga{}'s dynamic approach for information retrieval makes it adapt to different project contexts. Furthermore, all selected projects are hosted on GitHub, showing that our approach performs just as well with GitHub's issue-tracking system as it did with Jira in RQ1. \ga{}'s architecture is also easily extendable, enabling support for additional issue management platforms with minimal effort. The RQ3 dataset includes an equal number of issues from Python and Go projects, on which \ga{} achieves mean accuracies of 93\% and 85\%, respectively.

\begin{tcolorbox}
On 120 issues from post-training-cutoff real-world repositories, \ga{} achieves an average Hit@1 of 0.89, confirming strong generalizability to new, unseen data in both Python and Go projects.
\end{tcolorbox}

\subsection{\textbf{RQ4: What are the costs of the approach?}}
\label{sec:rq4}

In this section, we evaluate the operational cost of \ga{} in three different dimensions. This analysis provides a practical understanding of \ga{}'s cost-effectiveness.
\subsubsection{Experimental Setup.}
Our evaluation is based on the RQ1 experiments conducted using the EALink dataset.

\paragraph{\textbf{Evaluation Metrics.}}
We measure \ga{}s' cost from three perspectives: (i) the time to perform ILR for each issue, (ii) the number of tokens consumed by queries to the LLM (Chat-GPT4o-mini), and (iii) the monetary costs associated with token consumption, based on OpenAI's pricing as of May 2025. 

\subsubsection{Experimental Results.}
Our findings are summarized in \tabref{costs}. Based on the results, \ga{}'s average ILR time per issue is \checkNum{23} seconds. Additionally, the average number of tokens consumed by querying the LLM to resolve a single issue-commit link is approximately \checkNum{115k} tokens, which translates to \checkNum{0.01} US dollars. 

\change{To put these numbers into perspective, the EALink paper reported 14 hours of training and 126 seconds just to score 1,000 issue–commit pairs on the Isis project~\cite{zhang2023ealink}. EasyLink, on the other hand, requires as much as two terabytes of memory to handle repositories with more than 10,000 commits~\cite{huang2025back}, whereas \ga{} ran efficiently on a machine with only 16 GB of memory.}

\begin{tcolorbox}
\ga{} resolves each issue-commit link in under 24 seconds and \$0.01 on average, making it a cost-efficient alternative to prior DL methods that require extensive training resources.
\end{tcolorbox}

\section{Discussion and Future Work}\label{sec:disc}

\rev{\subsection{\ga{}'s Limitations}\label{sec:FailureAnalysis}
To better understand the limitations of \ga{} and the impact of LLM non-determinism, we conducted a manual qualitative analysis of cases where~\ga{} failed to identify the final resolving commit. Out of the 6,468 failures identified in the evaluation set, we randomly sampled 363 cases for detailed inspection, maintaining a 95\% confidence level with a 5\% margin of error. Each case was analyzed by two authors independently, with discrepancies resolved through consensus-based discussion. 
We observed three primary types of failures: \textit{Resource Exhaustion} (reaching iteration or context limits), \textit{Explicit Termination} (the agent invokes the \texttt{GiveUp} function), or \textit{Misidentification} (returning an incorrect commit hash). We identified several recurring root patterns that contribute to these errors regardless of the failure types. Note that a single failure may exhibit multiple patterns; therefore, the percentages below do not sum to 100\%.

\paragraph{\textbf{LLM Non-Determinism and Hallucinations (78\%).}}
The most prevalent source of failure stems from LLM-specific stochasticity and hallucinations. A frequent pattern observed is a lack of \textit{syntactic infidelity} when passing data between functions. For instance, when the LLM attempts to use a commit hash retrieved from one function as an argument for another (e.g., \texttt{commit\_diff}), it may slightly alter the string. Such non-deterministic mutations render the hash invalid, causing function errors. While \ga{} often attempts to self-correct in subsequent iterations, these cycles consume the finite iteration and token budget, leaving less room for substantive reasoning. Additionally, we observed instances where the LLM entered indefinite loops while paginating through exceptionally large commit lists (exceeding 2,000 commits). In these cases, the massive data volume caused the model to hallucinate incorrect pagination states and repeat the same API calls until the system resources are exhausted.

\paragraph{\textbf{Misleading or Insufficient Information (35\%).}} 
In many instances, the source data lacked sufficient signal for successful recovery. Specifically, 27\% of these cases featured vague issue titles without accompanying descriptions. In 8\% of cases, the issue description was obscured by external links inaccessible to the agent. While \ga{} could be extended to fetch such external content, providing this contextual expansion remains beyond the current scope of this work.

\paragraph{\textbf{Structural and Historical Irregularities (23\%).}} 
Certain repository structures presented challenges that \ga{} was not initially designed to handle. For example, we noted failures when a ``placeholder'' commit with no code changes and a minimal message (e.g., ``Fix [CALCITE-12345]'') followed the actual resolving commit chain. Due to the lack of contextual data in this final commit, \ga{} may disregard it in favor of more ``informative'' intermediate changes. Furthermore, 13\% of failures involved issues that were reopened after an initial closure. In 2\% of cases, resolutions spanned multiple branches; in these scenarios, the parental and temporal relationships between commits are no longer upheld, causing a violation of \ga{}'s single-branch assumption and leading to the identification of intermediate rather than final commits.

\paragraph{\textbf{Noisy Contextual Data (4\%).}} 
In a small subset of cases, high-volume query responses overwhelmed the agent's reasoning capacity. For example, querying a common file name like \texttt{map.java} produced exhaustive listings across the codebase. The resulting noise interferes with the LLM's ability to isolate relevant files, suggesting that more refined search heuristics could further improve the agent's precision in large-scale repositories.}

\subsection{\textbf{Key Insights and Implications}}
The experiments in this work encompassed a broad range of scenarios, yielding several actionable insights and recommendations for future work.

\paragraph{\textbf{Broader Applicability.}} 
In RQ1, we showed that giving the model access to a broader, but less frequently used, set of data sources can improve accuracy on projects where the usual sources are less informative. Although we designed \ga{} specifically for issue-to-commit link recovery (ILR), the same approach can also benefit other tasks, such as various forms of TLR.
A key factor in \ga{}'s consistent performance is its on-demand access to all relevant data sources. By letting the LLM pull exactly the information it needs, \ga{} adapts to different project contexts, handles corner cases, and avoids over-reliance on a fixed feature set. We recommend that researchers adopt a similar approach to improve coverage of corner cases and enhance overall robustness in other TLR tasks.

\paragraph{\textbf{Scalable Context Access.}} 
We designed two key features that enable \ga{} to access an effectively unlimited amount of data without exhausting the context window:
(1) \emph{Pagination:} Any function that returns a long list accepts a \code{pagination} parameter. This ensures each response stays within the token limit; and 
(2) \emph{Feedback pruning:} When a function returns a large output, the agent can trigger a feedback mechanism that retains only the data points deemed truly relevant to the ILR task by the LLM. This ensures that the overall chat history stays within the token limit.
Together, these strategies help \ga{} scale gracefully while maintaining focus, and we encourage future agent developers to adopt similar techniques for handling large outputs effectively.

\paragraph{\textbf{Balancing Determinism.}}
A critical aspect of agent development is deciding where to draw the line between traditional programming and delegation to the LLM. That is, determining which tasks should be baked into functions and which should be left to the LLM to handle. This represents a fundamental trade-off between determinism and agency that developers must carefully consider. Granting the LLM more autonomy can lead to more flexible and adaptive solutions, while increasing determinism through code can reduce errors but potentially limit the system's ability to handle edge cases. This trade-off can significantly impact overall performance. During the development of \ga{}, we observed a notable accuracy boost by introducing more deterministic components into the LLM's toolset. For instance, instead of relying on the LLM to accurately filter commit histories on its own, we provided dedicated commit-retrieval functions that query by specific attributes. By encoding such behaviors directly into the system, we eliminated a class of simple reasoning errors. However, excessive determinism can undermine one of the key strengths of agentic approaches, namely their adaptability. We therefore advocate for an incremental strategy: start with high agency by giving the LLM broad access to data, and gradually convert recurring mistakes or inefficiencies into deterministic functions as they emerge.

\paragraph{\textbf{Practical Efficiency.}} 
\ga{} interacts with LLMs solely through APIs, requiring no additional training or specialized hardware. It can be deployed in any environment and easily extended to incorporate domain-specific data sources. As a ready-to-use tool, \ga{} can be applied to issues and projects beyond those included in this study, as demonstrated in \secref{rq3}. We believe that agentic solutions like \ga{} hold great potential for further development, accelerating both innovation and research reuse. In contrast to traditional AI models, which often require custom feature engineering and retraining to handle new data, this level of flexibility provides a significant advantage. While traditional approaches may perform well in certain scenarios, agentic solutions can be far more practical in real-world settings. For example, as shown in \secref{rq3}, our traditional baseline EALink requires 14 hours of training and 126 seconds to score 1,000 issue-commit pairs in a particular project. In contrast, \ga{} takes just 29 seconds on average to retrieve the latest resolving commit for any given issue in the same project. Given the 29,101 commits in the target project, EALink's inference alone would take nearly an hour (29,101 × 0.126s = 3,667s), not including training time. Meanwhile, \ga{} achieves a Hit@1 score of 0.88 on this project, compared to EALink's 0.24, and requires no setup, training, or GPU infrastructure. Therefore, we advocate for agentic solutions to accelerate progress and improve accessibility in research.

\section{Threats to Validity}\label{sec:threats}
\paragraph{\textbf{Internal Validity.}}
We leveraged widely adopted open-source tools, such as the OpenAI \change{LLM} SDK, which are well-integrated within the open-source community. However, using these tools also carries the risk of inheriting any existing flaws they might have. We also recognize the possibility of hidden bugs in our implementation. To address these, we performed multiple manual tests, developed a comprehensive test suite, and made our code and data publicly available~\cite{Git-Anchor} for community inspection and feedback. 

\rev{\paragraph{\textbf{External Validity.}} The primary objective of this work was to introduce the dynamic heuristic search paradigm and validate its effectiveness in addressing the limitations of prior ILR approaches. To achieve this, we evaluated~\ga{} with a representative set of standard technologies, including GitHub and Jira for issue tracking, Tree-sitter for multi-language parsing, and ChatGPT-4o-mini as a cost-effective LLM. While this configuration validated the proposed paradigm, we recognize that a comprehensive benchmark across alternative settings, such as different LLMs or issue-tracking platforms, is valuable for establishing broader robustness. To facilitate such comparative analysis, we designed~\ga{} with a modular architecture that allows individual components (e.g., LLM model) to be easily substituted. We encourage future research to leverage this flexibility to benchmark the framework under diverse settings and uncover new insights.}

\paragraph{\textbf{Construct Validity.}} The timing results reported in our experiments are influenced by the performance of the underlying LLM API, in this case, OpenAI. Factors such as server congestion, rate limits, and the inherent randomness of LLM behavior can cause variability in response times. To minimize this variance, we repeated the experiments three times and reported the average. 

\paragraph{\textbf{Reproducibility.}} We provide the entire experiment, including the source code for \ga{} and scripts for data generation and evaluation online~\cite{Git-Anchor}.

\section{Related Work}\label{sec:related}
\paragraph{\textbf{Software Traceability Link Recovery (TLR).}}
\rev{These methods seek to recover semantic relationships across software artifacts, including requirements, architecture diagrams, and source code. Early work used IR techniques~\cite{heydarnoori2011two,falessi2017estimating,mills2018automatic,hayes2006advancing,antoniol2002recovering,gethers2011integrating,mahmoud2015role,rempel2013towards,delucia2004enhancing,marcus2003recovering,dekhtyar2007technique,asuncion2010software,tahmooresi2020studying,gao2023using,hey2021improving,mahmoud2013supporting,kuang2019using,rodriguez2020ir,rodriguez2020multiobjective}, while later studies adopted ML and DL methods in both supervised and semi-supervised settings to better capture artifact semantics~\cite{dong2022semisupervised,moran2020improving,guo2017semantically,lin2021traceability}. Despite their performance gains, these approaches often have high training costs. Several studies~\cite{keim2024recovering,nishikawa2015recovering,rodriguez2021leveraging} have exploited the relationships of intermediate artifacts with both ends of the link. These methods are ineffective when intermediate artifacts are unavailable or of low quality. Most recently, studies have leveraged transformer-based models~\cite{needle_in_haystack,silentVul,abedini2024can} and LLMs enhanced by prompt engineering~\cite{rodriguez2023prompts}, and RAG~\cite{fuchss2025lissa}. However, LLMs remain constrained by token limits, which often prevent them from processing lengthy but critical data, such as large codebases and documentation.}

\paragraph{\textbf{Issue-to-Commit Link Recovery.}} 
\rev{Existing ILR approaches primarily fall into two paradigms: ranking systems and classification models. Ranking systems typically assign a relevance score to individual issue-commit pairs based on textual feature similarity, enabling the retrieval of a top-$K$ list of most likely resolving commits. In Hybrid-Linker~\cite{mazrae2021automated}, Rostami et al. trained one model on commit and issue text, and another on code and structured data, achieving strong precision by combining their outputs. DeepLink~\cite{ruan2019deeplink} proposed a semantically enhanced method, employing RNNs to learn vector representations of issue and commit texts. BTLink~\cite{lan2023btlink} advanced this by using different pre-trained transformer models to learn programming language and natural language separately. Zhang et al.~\cite{zhang2023ealink} highlighted that issue-commit links are often one-to-many. To capture this characteristic, they constructed a dataset containing the full set of commits involved in each resolution and utilized it to train a model via knowledge distillation and multi-task contrastive learning. While this improves the recall of relevant commits, their inference remains limited to scoring individual issue-commit pairs and lacks reasoning over the aggregated changes to identify the final resolution point. EasyLink~\cite{huang2025back} avoids the costly training required by the aforementioned approaches. Instead, it applies an LLM to rerank candidate commits retrieved from a vector database, effectively bridging the semantic gap between issues and commits. However, EasyLink's initial retrieval stage relies purely on textual embedding similarity. This creates a potential bottleneck, as textual similarity between a commit and an issue does not necessarily imply a fixing relationship~\cite{huang2025back}. Furthermore, much like prior work, EasyLink is constrained by a narrow feature set and overlooks contextual signals such as issue comments and source code changes, which our approach demonstrates are vital for ILR effectiveness.

Classification approaches treat ILR as a binary decision problem, where the model assigns a direct binary label (Link/No-Link) to an issue-commit pair without generating a relevance score or ranked list. Recently, LLMs have been adapted for this task using a cloze (fill-in-the-blank) formulation. MPLinker~\cite{wang2025mplinker} and PromptLink~\cite{deng2024promptlink} adopt this strategy, utilizing multi-template prompt tuning to improve model generalization.

As summarized in \tabref{related_work}, both of the aforementioned ILR paradigms rely on evaluating issue-commit pairs in isolation. As discussed in Sections~\ref{sec:intro} and~\ref{sec:motiv}, this analysis suffers from several limitations. Notably, it fails to capture the full resolution logic and the cumulative nature of software fixes. To address these challenges, \ga{} represents a paradigm shift from isolated pair analysis to a dynamic heuristic search over the version control graph. This search aggregates the entire chain of related changes and identifies the final resolving commit as the definitive resolution point, ensuring that the complete fix history is recovered.
}

\paragraph{\textbf{LLM Agents.}}
\rev{In recent studies, LLM agents have enabled reasoning over complex codebases and achieved strong results in various software engineering tasks~\cite{jin2024llms,roychoudhury2025agentic,hassan2025agentic,li2025rise,liu2024large,xia2025demystifying}, including code localization~\cite{chen2025locagent}, fault localization~\cite{kang2023autofl}, automated program repair~\cite{bouzenia2024repairagent,yu2025patchagent}, automated issue resolution~\cite{yang2024swe,wang2024openhands,zhang2024autocoderover,li2025swe,ma2025alibaba}, and software testing~\cite{bouzenia2024younameit,cheng2025agentic}. To the best of our knowledge, however, no existing work has applied LLM agents to the TLR or ILR problems.}

\newcolumntype{C}[1]{>{\centering\arraybackslash}m{#1}}
\newcolumntype{L}[1]{>{\raggedright\arraybackslash}m{#1}}

\begin{table*}[t]
\centering
\rev{
\small
\renewcommand{\arraystretch}{1.3}
\setlength{\tabcolsep}{3pt}
\caption{\rev{Comparison of Prior Work Paradigms and \ga{}}}
\label{tab:related_work}
\vspace{-0.4cm}
\resizebox{\textwidth}{!}{%
\begin{tabular}{|
  C{2cm}|
  C{2.4cm}|
  L{4.0cm}|      
  C{1.35cm}|
  C{0.45cm}C{0.45cm}C{0.45cm}C{0.45cm}|
  C{0.45cm}C{0.45cm}C{0.45cm}C{0.45cm}C{0.45cm}|
  C{1cm}|
  C{1.8cm}|
  C{2.5cm}|
}
\hline

\multirow{5.5}{*}{\textbf{ILR Paradigm}} &
\multirow{5.5}{*}{\textbf{\makecell{Paradigm\\Description}}} &
\multicolumn{1}{C{4.0cm}|}{\multirow{5.5}{*}{\textbf{Study}}} &
\multirow{5.5}{*}{\textbf{\makecell{Needs\\Training?}}} &
\multicolumn{10}{c|}{\textbf{Contexts Used}} &
\multirow{5.5}{*}{\textbf{Feature Set}} &
\multirow{5.5}{*}{\textbf{Resolution Scope}} \\
\cline{5-14}

& & & &
\multicolumn{4}{c|}{\textbf{Issue}} &
\multicolumn{5}{c|}{\textbf{Commit}} &
\multirow[b]{2}{*}{\textbf{\makecell{Source\\Code}}} &
& \\
\cline{5-8} \cline{9-13}

& & & &
\rotatebox{90}{Title } &
\rotatebox{90}{Desc. } &
\rotatebox{90}{Disc. } &
\rotatebox{90}{Meta } &
\rotatebox{90}{Msg. } &
\rotatebox{90}{Diff } &
\rotatebox{90}{Files } &
\rotatebox{90}{Hist. } &
\rotatebox{90}{Meta } &
& & \\
\hline\hline

\multirow{5}{*}{\textbf{Ranking}} &
\multirow{5}{=}{Assigns relevance score to commits \& retrieves Top-$K$} &
Hybrid-Linker~\cite{mazrae2021automated}: \scriptsize \textit{Ensemble ML} & \cmark &
\cmark & \cmark & \xmark & \cmark &
\cmark & \cmark & \xmark & \xmark & \xmark &
\xmark &
\multirow{6}{=}{\centering Static} &
\multirow{6}{=}{\centering Isolated Issue-Commit Pairs} \\

& & DeepLink~\cite{ruan2019deeplink}: \scriptsize \textit{DL (RNN)} & \cmark &
\cmark & \cmark & \xmark & \xmark &
\cmark & \cmark & \xmark & \xmark & \xmark &
\xmark &
& \\

& & BTLink~\cite{lan2023btlink}: \scriptsize \textit{PTMs} & \cmark &
\cmark & \cmark & \xmark & \xmark &
\cmark & \cmark & \xmark & \xmark & \xmark &
\xmark &
& \\

& & EALink~\cite{zhang2023ealink}: \scriptsize \textit{Distilled PTM} & \cmark &
\cmark & \cmark & \xmark & \xmark &
\cmark & \cmark & \cmark & \xmark & \xmark &
\xmark &
& \\

& & EasyLink~\cite{huang2025back}: \scriptsize \textit{IR + LLM Re-ranking} & \xmark &
\cmark & \cmark & \xmark & \xmark &
\cmark & \xmark & \xmark & \xmark & \xmark &
\xmark &
& \\

\cline{1-14}

\textbf{Classification} &
Assigns binary label (link/no-link) &
MPLinker~\cite{wang2025mplinker} \& PromptLink~\cite{deng2024promptlink}: \scriptsize \textit{Cloze Task} & \cmark &
\xmark & \cmark & \xmark & \xmark &
\cmark & \cmark & \xmark & \xmark & \xmark &
\xmark &
& \\
\hline

\textbf{Dynamic Search} &
Searches with aggregated reasoning of relevant commits &
\textbf{\ga{}}: \scriptsize \textit{Agentic LLM} & \xmark &
\cmark & \cmark & \cmark & \cmark &
\cmark & \cmark & \cmark & \cmark & \cmark &
\cmark &
Dynamic \scriptsize (via LLM reasoning) &
Reason over combined changes of multiple commits \\
\hline

\end{tabular}%
}
}
\end{table*}

\section{Conclusions}\label{sec:conclusion}



In this paper, we presented \ga{}, the first LLM-agent-based approach specifically developed for ILR. \ga{} utilizes a state-of-the-art LLM (i.e., ChatGPT-4o-mini) without the need for additional training or fine-tuning. Its agent-based architecture enables access to the full problem context, including the entire codebase and the complete issue discussion history. Furthermore, \ga{} is language-agnostic by design. We evaluated \ga{} on six widely used open-source Apache projects and compared its performance to state-of-the-art baselines. \ga{} consistently outperformed all prior methods, with improvements ranging from 41\% to 714\%. These results highlight \ga{}'s ability to dynamically access and leverage contextual information that previous methods could not effectively utilize. \change{Furthermore, to better understand the contribution of each newly introduced data source, we conducted an ablation study. The findings show that issue comments boost performance by about 17\%, while the codebase contributes a smaller but still meaningful 4\% gain. The ablation also demonstrated \ga{}'s flexibility: it can selectively draw on different sources depending on the project context, allowing it to remain effective even when certain information (e.g., issue description or commit messages) is missing or sparse. This underscores the advantages of an agent-based design, allowing \ga{} to address a broad range of real-world scenarios more effectively than static, text-only methods.}

To further assess its generalizability, we evaluated \ga{} on 120 randomly selected issues drawn from six additional GitHub repositories that were unseen during development. \ga{} successfully recovered links for \checkNum{107} of these issues, corresponding to an accuracy of \checkNum{89}\%. These results highlight \ga{}'s adaptability and robustness across heterogeneous codebases. Finally, our cost analysis shows that, for a given issue, \ga{} completes ILR in under 24 seconds at an average cost of \$0.01, indicating the cost-effectiveness of the approach. Moreover, \ga{} is designed with extensibility and seamless integration in mind. New data-extractor functions can be incorporated with minimal effort, and substituting a next-generation LLM requires only redirecting the agent to a new API endpoint. This plug-and-play flexibility ensures that \ga{} can evolve in step with advances in LLM technology and adapt readily to emerging environments.

\section{Data Availability}
Our replication package~\cite{Git-Anchor} contains the code, data, and all necessary materials to reproduce the results presented in this paper.


\bibliographystyle{ACM-Reference-Format}
\bibliography{references}

\end{document}